\documentclass{aa}  
\usepackage{graphicx}	% Including figure files
\usepackage{amsmath}	% Advanced maths commands
\usepackage{amsmath, amsfonts, amssymb, graphics,graphicx, wrapfig, nicefrac}
\usepackage{multirow}
\usepackage{multicol}
\usepackage{graphicx}
\usepackage{booktabs}
\usepackage{rotating}
\usepackage{mathrsfs}
\usepackage{float}
\usepackage[version=3]{mhchem} % Formula subscripts using \ce{}
\graphicspath{{./}}
\usepackage{txfonts}
\usepackage[colorlinks=true,linkcolor=blue, citecolor=blue,urlcolor=magenta]{hyperref}
\begin{document}

   \title{A HINSA view of cosmic-ray ionization in IC~348 and NGC~1333: evidence for a strong low-energy cosmic-ray disparity}
   \titlerunning{The cosmic-ray ionization rate disparity between IC~348 and NGC~1333}

   \author{Gan Luo
          \inst{1}
          \and
          Marco Padovani\inst{2}
          \and 
          Daniele Galli\inst{2}
          \and 
          Thomas G. Bisbas\inst{3}
          \and 
          Brandt A. L. Gaches\inst{4}
          \and 
          Di Li \inst{5,6}
          \and
          Marko Kr{\v{c}}o \inst{6}
          \and 
          Ningyu Tang \inst{7}
          }

   \institute{Institut de Radioastronomie Millimétrique, 300 rue de la Piscine, 38400, Saint-Martin d’Hères, France\\
            \email{luo@iram.fr}
        \and
            INAF-Osservatorio Astrofisico di Arcetri, Largo E. Fermi 5, 50125 Firenze, Italy
        \and 
            Research Center for Computational Earth and Space Science, Zhejiang Lab, Hangzhou, 311100, PR China
        \and 
            Faculty of Physics, University of Duisburg-Essen, Lotharstraße 1, 47057 Duisburg, Germany
        \and
            New Cornerstone Science Laboratory, Department of Astronomy, Tsinghua University, Beijing 100084, China
        \and
            State Key Laboratory of Radio Astronomy and Technology, National Astronomical Observatories, Chinese Academy of Sciences, Beijing 100101, China
        \and
            Department of Physics, Anhui Normal University, Wuhu, Anhui 241002, China
             }

   \date{Received xx; accepted xx}

\abstract{The cosmic-ray ionization rate (CRIR) is one of the fundamental parameters influencing the chemical and dynamical evolution of molecular clouds. Although observations in recent years have revealed high CRIR values in massive star-forming regions and in the vicinity of protostars, the sources and acceleration mechanisms of cosmic rays remain uncertain. In this work, we present our new estimates of CRIR using the H\,{\sc i} narrow self-absorption (HINSA) technique towards two nearby low-mass star-forming clouds, IC~348 and NGC~1333. In both clouds, the CRIR decreases with increasing H$_2$ column density, but IC~348 exhibits values that are roughly an order of magnitude higher than those in NGC~1333. To interpret this contrast, we model the low-energy spectrum of CRs in a finite slab attenuation framework, using additional constraints from the high-energy CR spectrum inferred from Fermi $\gamma$-ray observations. The best-fit spectra reproduce the observed CRIR profiles and the contrast between IC~348 and NGC~1333 suggests an order of magnitude difference in low-energy CR populations, likely originating from local acceleration sources beyond protostars (e.g., stellar-wind termination shocks), and partly from the same sources responsible for the GeV $\gamma$-ray excess. Although uncertainties in cloud structure and gas density may affect the absolute CRIR values, they do not erase the pronounced disparity between the two regions.}


\maketitle
\nolinenumbers
%
%________________________________________________________________

\section{Introduction}

Low-energy cosmic rays (LECRs, e.g., $E<1$\,GeV) play a crucial role in the physical and chemical evolution of the interstellar medium. They can penetrate deep into molecular cloud clumps, ionizing molecular hydrogen (H$_2$) and initiating chemical reactions that produce key ions \citep[e.g., H$_3^+$,][]{Dalgarno2006,Grenier2015,Gaches2026}. The resulting ionization fraction governs the coupling between magnetic fields and gas, thereby regulating the dynamics during star formation processes \citep{Padovani2020,Gabici2022}. While LECRs are thought to be accelerated by physical processes such as the first-order Fermi acceleration, their physical origins and their feedback on star formation remain poorly constrained \citep{Drury1994,Blasi2013}.

When Galactic LECRs propagate into dense molecular clouds, they lose energy due to processes such as ionization and bremsstrahlung \citep{Padovani2020}. Measured values of the cosmic-ray ionization rate (CRIR) $\zeta^{\rm ion}$, defined here as the ionization rate per H$_2$ molecule, decrease from low to high column densities \citep{Sabatini2023,Bialy+2026,Indriolo2026,Neufeld+2026}, consistent with CR attenuation models without internal acceleration sources \citep{Padovani2018a}.

Although the discrepancy between the CRIR inferred from {\it Voyager} measurements and that of nearby molecular clouds has been debated for a decade \citep{Neufeld2024}, growing evidence suggests an environmental dependence of the CRIR \citep{Phan2023,Redaelli2025}. This is particularly evident in the anomalously high values observed in massive star-forming regions and protoclusters \citep{Ceccarelli2014,Fontani2017,Luo2024b,Tang2026}. These observations suggest an {\it in situ} origin of LECRs that is most likely linked to star formation activities \citep[e.g., protostellar jet shocks,][]{Padovani2015,Padovani2016,Gaches2018,Gaches2019}. However, the CR sources and acceleration mechanisms responsible for the high values remain uncertain.

The situation becomes increasingly complex when comparing recent CRIR mapping inferred from molecular line observations with that from diffuse sub-GeV $\gamma$-ray observations in the nearby low-mass star-forming cloud NGC~1333. While the former indicate an elevated CRIR in the vicinity of embedded young stellar objects (YSOs) \citep{Pineda2024}, the latter reveal a deficit in the diffuse $\gamma$-ray residual within NGC~1333 \citep{Yang2023}. Conversely, the neighboring cloud IC~348 exhibits a positive $\gamma$-ray residual and a significantly harder energy spectrum of CRs \citep{Yang2023}, despite having lower star-formation efficiency and YSO density compared to NGC~1333 \citep{Young2015}. Consequently, it remains unclear whether star-formation activity is the driver of these discrepancies, and the extent to which the local CRIR in NGC~1333 overwhelms the Galactic CR background.

In this paper, we present new estimates of CRIR toward IC~348 and NGC~1333 using the H\,{\sc i} narrow self-absorption (HINSA) technique initially presented by \citet{Li2003} and \citet{GoldsmithLi2005}. Our observations reveal a clear difference in CRIR between IC~348 and NGC~1333, where the former is almost an order of magnitude higher than the latter, although the latter shows more intense star formation activity. 
The observations and data used in this work are described in Sect.~\ref{sec:obs}. We present the calculation of CRIR and the modeling of the LECRs spectrum in Sect.~\ref{sec:results}. We discuss the uncertainties and the possible origin of the observed disparity in Sect.~\ref{sec:discussion}. The conclusion is presented in Sect.~\ref{sec:conclusion}.

\section{Observations}\label{sec:obs}

\subsection{H\,{\sc i} observations}\label{sec:fast_hi}

The H\,{\sc i} observations toward NGC~1333 were taken with the 19-beam receiver of the Five-hundred-meter Aperture
Spherical radio Telescope \citep[FAST,][]{Nan2011} using ``DriftWithAngle'' observing mode from Oct. 02 to Dec. 04, 2022 (Project ID: PT2022\_0145, PI: Gan Luo), as part of the Commensal Radio Astronomy FAST Survey \citep[CRAFTS,][]{Li2018}.  
The calibration of the raw data was performed with a novel high-cadence-CAL technique (Kr{\v{c}}o et al. in prep).  The calibrated data were regridded to a pixel size of 1.5$'$ and a spectral resolution of 0.2~km~s$^{-1}$, which are publicly available through the CRAFTS database\footnote{https://doi.org/10.57760/sciencedb.07779.} \citep{Li2024}. The reached rms is $\sim$0.1~K per pixel per channel. 
The H\,{\sc i} data toward IC~348 were extracted from the Galactic Arecibo L-Band Feed Array H\,{\sc i} (GALFA-H\,{\sc i}) survey \citep{Peek2018}. The spectral resolution is $\sim$0.18~km~s$^{-1}$ and the noise is 80~mK per 1~km~s$^{-1}$.

\subsection{Archival $^{13}$CO data}\label{sec:co data}

We used $^{13}$CO $J=1$--0 observations from the COMPLETE survey \citep{Ridge2006}. 
The beam size is $46''$, and the data were regridded to $\sim23''$, resulting in rms noise levels of 0.08~K per 0.066~km~s$^{-1}$ in the two regions. The main beam efficiency is 0.49 at 110~GHz.
The $^{13}$CO $J$=3--2 observations were obtained with the HRAP imaging array in JCMT \citep{Curtis2010a}. The beam size is 17.7$''$, and the beam efficiency is 0.66 at 330~GHz. 
The resultant rms noise level is 0.7~K per 0.05~km~s$^{-1}$ channel. Both $^{13}$CO $J$=1--0 and 2--1 data were convolved to the same angular resolution of H\,{\sc i} and regridded to the same pixel size of H\,{\sc i} images for joint analysis.

\subsection{Archival continuum data}\label{sec:obs_continuum}

We used the maps of dust temperature $T_{\rm d}$ and H$_2$ column density $N({\rm H}_2)$ from the {\it Herschel} Gould Belt Survey \citep[HGBS;][]{Andre2010,Pezzuto2021}. We consider the $N({\rm H}_2)$ values to have an uncertainty of 12\% \citep{Roy2014}. The final products have an angular resolution of 36.3$''$. We performed the same convolution and regridding procedures as for $^{13}$CO on these data.

\section{Results}\label{sec:results}

\subsection{Abundance of HINSA}\label{sec:hinsa}

The column density of atomic hydrogen, $N({\rm HINSA})$ is given by \citep{Li2003}
\begin{equation}
    N({\rm HINSA}) = 1.95\times10^{18} \tau \ T_{\rm k} \Delta V \; {\rm cm^{-2}},
\end{equation}
where $\tau$ is the optical depth of HINSA, $\Delta V$ is the linewidth, and the kinetic temperature of H\,{\sc i}, $T_{\rm k}$ is assumed to be equal to $T_{\rm d}$. We used the second-order derivative method developed by \citep{Krco2008} to obtain $\tau$ and $\Delta V$ of HINSA.

The HINSA fitting follows the similar procedures as described in \citet{Tang2020} and \citet{Luo2025}. The velocity of $^{13}$CO is used as a prior parameter for HINSA fitting; the allowance deviation between HINSA and $^{13}$CO velocity is within 0.5~km~s$^{-1}$. 
We consider an additional 20\% uncertainty in $N({\rm HINSA})$ based on previous estimates \citep{Luo2024a}. To eliminate the contribution of photodissociation to $N({\rm HINSA})$, we only consider regions with $N({\rm H_2})>3\times10^{21}$~cm$^{-2}$. An example of the HINSA fitting is shown in Fig.~\ref{fig:hinsa_fit}.

\begin{figure*}
%\centering
\sidecaption
\includegraphics[width=12cm]{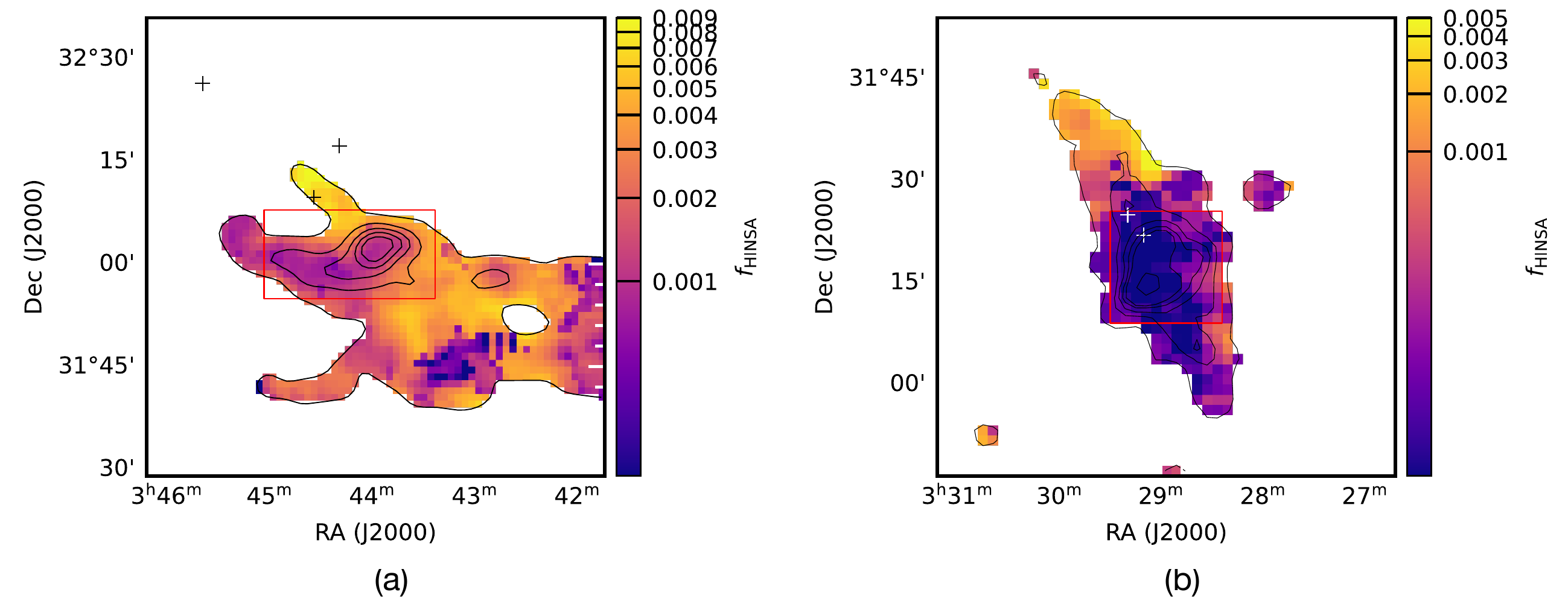}
\caption{{\it Panel~(a)\/}: color map of the HINSA fraction $f_{\rm HINSA}$ toward IC~348. Contours represent $N({\rm H_2})$ at $3\times10^{21}$~cm$^{-2}$ $\times$(1,2,3,4,5,10). Crosses mark the positions of B stars. The red rectangle outlines the region mapped in $^{13}$CO (3--2). {\it Panel~(b)\/}: same as Panel~(a), for NGC~1333. \label{fig:fig1}}
\end{figure*}

Figure \ref{fig:fig1}(a) and (b) show the spatial distribution of the fraction of HINSA relative to molecular 
hydrogen, $f_{\rm HINSA} = N({\rm HINSA})/N({\rm H_2})$, overlaid to $N({\rm H_2})$ contours toward IC~348 and NGC~1333, respectively. In both regions, $f_{\rm HINSA}$ is higher at the edge of the cloud and lower toward higher column densities. However, the average $f_{\rm HINSA}$ in IC~348, $(2.6\pm1.7)\times10^{-3}$, is much higher than in NGC~1333, $(4.7\pm8.6)\times10^{-4}$. 

\subsection{The gas volume density through {\sc radex} modeling}\label{sec:density}

The calculation of $\zeta^{\rm ion}$ requires a reasonable estimate of the gas volume density $n({\rm H_2})$. In this work, we used $^{13}$CO emission to derive $n({\rm H_2})$, as the HINSA feature is expected to trace envelopes where densities are comparable to $^{13}$CO \citep{Li2003,Luo2025}. We use the non-LTE radiative transfer code {\sc radex} \citep{Van2007} to constrain $n({\rm H_2})$. The collision rate coefficients are adopted from \citet{Yang2010}. We assume that the kinetic temperature $T_{\rm k}$ is the same as $T_{\rm d}$. 
The input free parameters in {\sc radex} modeling are $n({\rm H_2})$ and the $^{13}$CO column density $N({\rm ^{13}CO})$. We use the Markov Chain Monte Carlo (MCMC) method within the $emcee$ code \citep{Foreman-Mackey2013} to sample the posterior probability distributions of the above free parameters, in which the likelihood function of the posterior probability function is defined as:
\begin{equation}
\ln \ p = -\frac{1}{2} \sum_i \left [ \frac{\left ( {\it T}_{\rm obs}^i - {\it T}_{\rm mod}^i \right )^2}{{\sigma^i_{\rm obs}}^2} + \ln \left ( 2\pi {\sigma^i_{\rm obs}}^2 \right ) \right ],
\end{equation}
where the $T_{\rm obs}^i$ and $\sigma^i_{\rm obs}$ are the observed brightness temperature and its uncertainty of the $i$-th transition, respectively. $T_{\rm mod}^i$ is the modeled brightness temperature. 

The derived $n({\rm H_2})$ ranges from $(1.6\pm2.3)\times10^2$ to $(4.9\pm3.3)\times10^5$~cm$^{-3}$ in IC~348 and from $(1.1\pm0.2)\times10^3$ to $(2.2\pm0.2)\times10^4$~cm$^{-3}$ in NGC~1333. The median values of $n({\rm H_2})$ are $2.6\times10^3$ and $4.5\times10^3$~cm$^{-3}$ toward IC~348 and NGC~1333, respectively. We note that the assumption of $T_{\rm k} = T_{\rm d}$ is reasonable only if gas and dust are thermally coupled \citep[$n({\rm H}_2)\geq 10^{4.5}~{\rm cm^{-3}}$,][]{Goldsmith2001}, otherwise, the assumed $T_{\rm k}$ might be underestimated. We will discuss this bias in Sect.~\ref{sec:uncertainty of nh2}.

\subsection{The cosmic-ray ionization rate $\zeta^{\rm ion}$}
\label{sec:crir}

The abundance of molecular hydrogen in the dense, cold regions of a molecular cloud is determined 
by the balance of production of H$_2$ on the surface of 
dust grains at a rate $R_{\rm gr}n_{\rm H}n({\rm H})$, where $n_{\rm H}=n({\rm H})+2n({\rm H}_2)$ and $R_{\rm gr}$ is H$_2$ formation rate coefficient, and destruction of H$_2$ by cosmic-ray nuclei and secondary electrons in H$_2$ at a rate $\zeta^{\rm d}n({\rm H}_2)$.
At steady state,
\begin{equation}
    \zeta^{\rm d}n({\rm H}_2)=R_{\rm gr}n_{\rm H}n({\rm H}).
    \label{hbalance}
\end{equation}
Cosmic rays destroy H$_2$ by ionization and dissociation. Both processes are caused by direct collisions 
of cosmic-ray particles with H$_2$ molecules, with rates $\zeta^{\rm ion}$ and $\zeta^{\rm diss}$, respectively, and, indirectly, by the ultraviolet 
emission induced by cosmic-ray excitation of electronic states of 
H$_2$, with rates $\zeta^{\rm cr-uv, ion}$ and $\zeta^{\rm cr-uv, diss}$, respectively. The total rate of destruction of H$_2$ in Eq.~(\ref{hbalance}) is therefore 
\begin{equation}
    \zeta^{\rm d}=\zeta^{\rm ion}+\zeta^{\rm diss}+\zeta^{\rm cr-uv,ion}+\zeta^{\rm cr-uv,diss}.
    \label{zhtot}
\end{equation}

The derivation of the H$_2$ dissociation rate $\zeta^{\rm diss}$ by \cite{Padovani2018b} is updated in Appendix~\ref{revised_zetah}, including the contribution of dissociative ionization and, 
for cosmic-ray nuclei, dissociative electron capture. The result is $\zeta^{\rm diss}=0.589~\zeta^{\rm ion}$ 
at $N({\rm H}_2)=10^{22}$~cm$^{-2}$, weakly dependent on the 
cloud's column density and the assumed cosmic-ray spectrum. 
The rates of cosmic-ray induced photoionization and photodissociation of H$_2$,
calculated by \cite{Heays2017}, \cite{Padovani2024b} and \cite{Sipila2026}, are $\zeta^{\rm cr-uv,ion}=0.0545~\zeta^{\rm ion}$, $\zeta^{\rm cr-uv,diss}=0.831~\zeta^{\rm ion}$, respectively.
In summary,
\begin{equation}
\zeta^{\rm d} = k~\zeta^{\rm ion},
\end{equation}
with $k=1+0.589+0.0545+0.831=2.47$.

As for the H$_2$ formation rate coefficient on dust grains, 
the value of $R_{\rm gr}$ in Eq.~(\ref{hbalance}) is not well constrained.  
In the diffuse ISM, a value of $R_{\rm gr}= 3$--$4\times
10^{-17}$~cm$^3$~s$^{-1}$ has been derived from observations of ultraviolet
absorption lines with the Copernicus and Fuse satellites \citep{Jura75,Gry02}. 
In cold, dense gas with $T=10$~K, \citet{GoldsmithLi2005} estimated a fiducial value of $R_{\rm gr}=1.2\times 10^{-17}$~cm$^3$~s$^{-1}$. However, the actual grain size distribution in dense cloud, as well as the effects of ice coating of
grains on the sticking coefficient \citep[assumed constant and equal to $0.3$ by][]{GoldsmithLi2005} 
are uncertain. Recent modelings of H$_2$ formation considering both Langmuir-Hinshelwood and Eley-Ridel mechanisms on dust grain in various astrophysical environments show that $R_{\rm gr}$ can exceed $2\times10^{-17}$~cm$^3$~s$^{-1}$ in well-shielded dense environments \citep{Bron2014,Thi2018}. In dense photodissociation regions (PDRs), $R_{\rm gr}$ can be even higher due to the existence of small grains/PAHs and temperature fluctuation \citep{Wakelam2017}. Thus, we adopt $R_{\rm gr}=3\times10^{-17}$~cm$^{-3}$~s$^{-1}$ in our calculation.

The balance equation (\ref{hbalance}) now allows to 
obtain $\zeta^{\rm ion}$ as
\begin{equation}
    \zeta^{\rm ion}=\frac{\zeta^{\rm d}}{k}=\frac{R_{\rm gr}n_{\rm H}n({\rm H})}{k n({\rm H}_2)}
    \approx \frac{R_{\rm gr}n_{\rm H}}{k}f_{\rm HINSA},
    \label{eq:zeta_ion}
\end{equation}
where $f_{\rm HINSA}=N({\rm HINSA})/N({\rm H}_2)\approx n({\rm H})/n({\rm H}_2)$.

\begin{figure}
\centering
\includegraphics[width=0.95\linewidth]{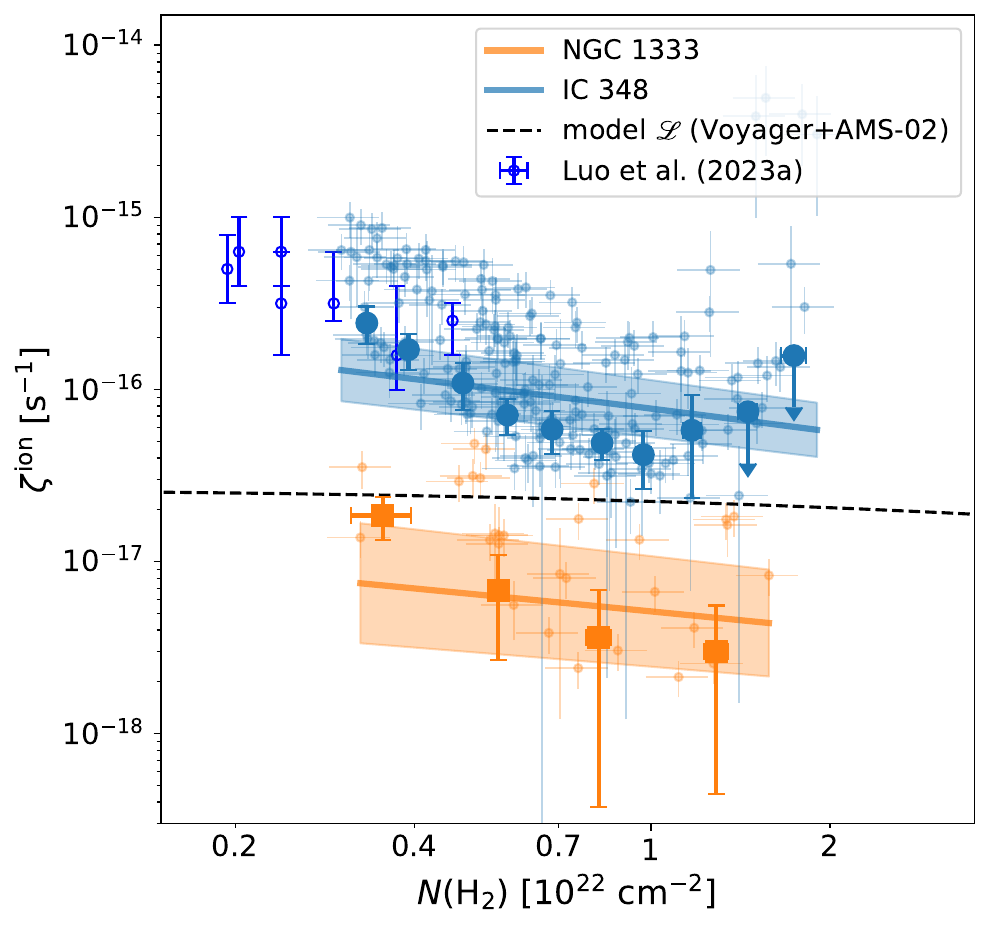}
\caption{Distribution of $\zeta^{\rm ion}$ as a function of column density $N({\rm H}_2)$ toward IC~348 (blue) and NGC~1333 (orange). The prediction of the theoretical proton spectrum from Voyager + AMS-02 data (model $\mathscr{L}$ from \citealt{Padovani2024b}) is overlaid. Black dots and red squares with error bars represent the weighted-average values in each $N({\rm H}_2)$ bin. Blue and orange lines with shadows represent modeling from the inferred CR spectrum (see Sect.~\ref{sec:modelofcrir}).
Dark blue empty circles show the estimates in IC~348 by \citet{Luo2023a}.
\label{fig:zvsN}}
\end{figure}

Figure \ref{fig:zvsN} presents the derived values of $\zeta^{\rm ion}$ for both clouds. It also shows binned data whose width was obtained by using the Freedman-Diaconis rule \citep{FreedmanDiaconis1981}. 
The expected trend of the CRIR for a proton spectrum based on Voyager \citep{Cummings+2016,Stone+2019} and AMS-02 \citep{Aguilar+2015}, previously referred to as the 
model $\mathscr{L}$ (low), is also displayed for reference. 
Other generic trends (e.g., models $\mathscr{H}$ and $\mathscr{U}$, see \citealt{Padovani+2022}) are deliberately not shown, 
as we are interested in models developed specifically for individual star-forming regions. 

In IC~348, the weighted-average $\zeta^{\rm ion}$ values range from $(5.35\pm1.49)\times10^{-17}$~s$^{-1}$ to $(2.33\pm0.55)\times10^{-16}$~s$^{-1}$. The last two upper limits at the highest $N({\rm H}_2)$ are not included, as the high values are dominated by the large volume density jump and are less reliable (see Sect.~\ref{sec:uncertainty of nh2} for more discussion). 
Our derived values and the decreasing trend for IC~348 are consistent with those from PDR modeling using various molecular line tracers at a similar spatial resolution \citep{Luo2023a}.

The weighted-average $\zeta^{\rm ion}$ values in NGC~1333 range from $(3.94\pm2.46)\times10^{-18}$~s$^{-1}$ to $(1.68\pm0.25)\times10^{-17}$ s$^{-1}$, an order of magnitude lower than those in IC~348. The CRIR values in both IC~348 and NGC~1333 show a decreasing trend as $N({\rm H}_2)$. 
Comparing our results in NGC~1333 with previous mapping studies, the values are reasonably matched at $N_{\rm H_2}\lesssim10^{22}$~cm$^{-2}$ \citep[$\zeta^{\rm ion}=(1.0\pm1.2)\times10^{-17}$~s$^{-1}$ in][]{Pineda2024}. Although \citet{Pineda2024} shows a rapid increase in $\zeta^{\rm ion}$ at higher column densities, the two methods differ substantially in angular resolution and tracers, and we do not have measurements at higher column densities; thus, they may be probing distinct gas components. In addition, the volume density estimates employed in their work is proportional to $N({\rm H}_2)$, which leads to a increasing trend of CRIR with column density. We discuss this further in Sect.~\ref{sec:interpretation}. 

\subsection{Modeling of low-energy spectrum of CRs}\label{sec:modelofcrir}

We introduce a new method for comparing the cosmic-ray ionization rate, $\zeta^{\rm ion}$ predicted by models with that estimated from observations. Instead of a semi-infinite 1D slab \citep{Padovani2009}, we consider a 1D slab with a finite total hydrogen column density $N_{\rm c}$.
The ionization rate at a given column density, $N=N({\rm H})+2N({\rm H}_2)$, is due to the flux $j_p(E,N)$ of cosmic-ray protons that has passed through a column density $N$ and to the cosmic-ray flux $j_p(E,N_{\rm c}-N)$ attenuated by $N_{\rm c}-N$, that is
\begin{align}\label{eq:zetadef}
\zeta^{\rm ion}(N) =& 2\pi \eta(1+\Phi) \int_I^\infty [\langle j_p(E,N)\rangle  \\
&+\langle j_p(E,N_{\rm c}-N)\rangle]\ \sigma^{\rm ion}_{p,\ce{H2}}(E)\ dE\,,\nonumber
\end{align}
where 
$I=15.44$~eV is the ionization threshold for \ce{H2} ionization, 
$\eta=1.51$ and $\Phi=0.73$ are the correction for the ionization by CR nuclei \citep{Padovani2009} and by secondary electrons \citep{Padovani+2026},
respectively,
and $\sigma^{\rm ion}_{p,\ce{H2}}(E)$ is the ionization cross section of molecular hydrogen by proton impact \citep{Rudd+1985} including the relativistic correction
\citep{Krause+2015}. 
Here, $\langle\cdot\rangle$ represents the flux averaged over the pitch angle \citep{Padovani2018a}
\begin{equation}
\langle j_p(E,N)\rangle = N\int_{N}^\infty\frac{j_p(E,\widetilde{N})}{\widetilde{N}^2}d\widetilde{N}\,.
\end{equation}
The most natural approach to comparing observations and models is to associate the ionization rate estimated from the observations 
to the value predicted by the model 
at the center of the cloud, that is, $N=N_{\rm c}/2$. Eq.~(\ref{eq:zetadef}) then reduces to
\begin{equation}\label{eq:zeta2}
\zeta^{\rm ion}(N_{\rm c}/2) = 4\pi \eta(1+\Phi) \int_I^\infty \langle j_p(E,N_{\rm c}/2)\rangle\ \sigma^{\rm ion}_{p,\ce{H2}}(E)\ dE\,.
\end{equation}

As for the parametrization of the proton spectrum incident on the cloud that best reproduces the observational data, we adopt the expression
\begin{equation}\label{eq:jparam}
j_p(p) = C\frac{p^\alpha}{(p+p_0)^\beta}\,,
\end{equation}
where $p$ is the particle's momentum.
This parametrization differs from that used in earlier works \citep[see, e.g.,][]{Ivlev2015,Padovani2018a} in two main respects: ({\it i}\/) it is expressed as a function of momentum rather than energy 
to avoid divergences in $\zeta^{\rm ion}$ when extrapolating to low energies, and ({\it ii}\/) it no longer relies on the Voyager data, since an increasing amount of observational studies 
seems to confirm that the flux of low-energy cosmic rays measured by these probes is not a good representation of the Galactic flux \citep[see, e.g.,][]{Redaelli2025,Indriolo2026,Bialy+2026,Neufeld+2026}. 
For this reason, the idea that each source should be modeled using a tailored local spectrum that differs from other regions of the sky because of the presence of nearby cosmic-ray sources such as supernova remnants and OB stars \citep{Aharonian+2019,Meyer+2024} is emerging.

\begin{figure}
\centering
\includegraphics[width=0.95\linewidth]{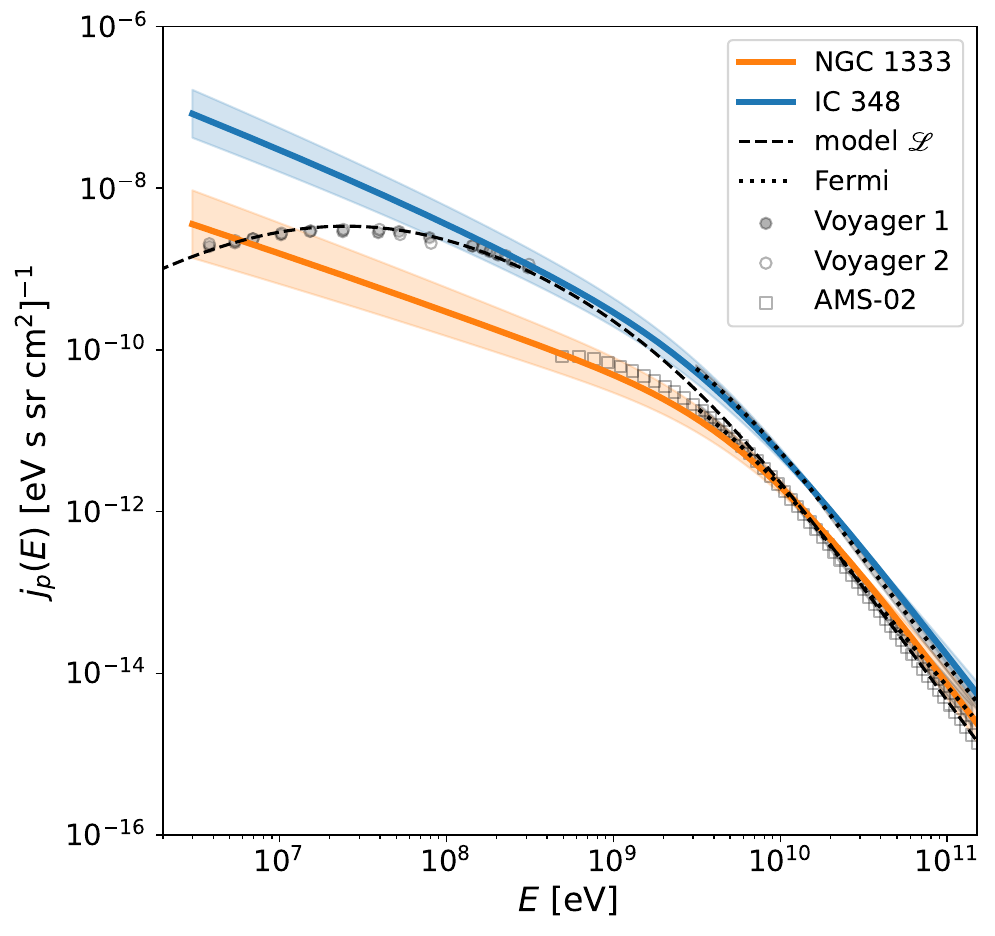}
\caption{CR proton spectra as a function of energy obtained with the best-fitting parameters in Table~\ref{tab:params} for NGC~1333 (orange solid line) and IC~348
(blue solid line). The shaded areas show the $\pm 1\sigma$ uncertainty. For comparison, the prediction for Voyager + AMS-02 data (model $\mathscr{L}$, dashed black line) is also shown.
Observational data from Fermi (dotted black line; \citealt{Jiang2025}),
Voyager 1 and 2 (empty and solid gray dots; \citealt{Cummings+2016,Stone+2019}) and AMS-02 (empty gray squares; \citealt{Aguilar+2015}).
The inset shows a zoom-in of the high energy region.
\label{fig:CRspectra}}
\end{figure}

For the two low-mass star-forming clouds examined in our study, the incident cosmic-ray spectra were derived from Fermi observations.
They exhibit different energy slopes in the energy range 3$-$100~GeV, and the spectrum
of IC 348 is 2$-$3 times higher than that of NGC 1333 \citep{Jiang2025}, confirming the variability of the cosmic-ray spectrum.
\citet{Jiang2025} derived a spectrum of cosmic-ray nuclei; therefore, to obtain the proton component, we divide it by a factor of 1.9 \citep{Kachelriess+2014}.
The proton spectra constrain the normalization constant, $C$, and the slope at high momentum, $\beta-\alpha$, in Eq.~(\ref{eq:jparam}), accounting for the fact that the
data have a $1\sigma$ uncertainty of 30\%.
We then vary the slope at low energies, $\alpha$, and the momentum at which the two slopes connect, $p_0$, propagate the cosmic-ray spectra using the 
continuous slowing-down approximation \citep{Takayanagi1973,Padovani2009}, and calculate $\zeta^{\rm ion}$ (Eq.~\ref{eq:zeta2}). Using a chi-squared test, we
determine the values of $\alpha$ and $p_0$ that best reproduce the observational estimates of $\zeta^{\rm ion}$. 
Table~\ref{tab:params} lists the best-fit parameters 
together with the best fit in momentum for the Voyager + AMS-02 data (model $\mathscr{L}$)
and Fig.~\ref{fig:CRspectra} shows the corresponding CR proton spectra as a function of kinetic energy. The CR proton toward NGC~1333 shows a deficit at ~GeV level compared to AMS-02 data, which is mainly due to slow diffusion processes \citep{Yang2023,Jiang2025}.
The minimum column density at which $\zeta^{\rm ion}$ is estimated from observations is $N(\ce{H2})\sim3\times10^{21}$~cm$^{-2}$. This is the stopping range for a proton
of $\sim3$~MeV \citep{Padovani2018a}; therefore, the modeled CR proton spectra cannot be extended below this energy threshold.
We also calculated the energy densities for the best-fit spectra of the two star-forming clouds, finding 0.34~eV~cm$^{-3}$ and 1.12~eV~cm$^{-3}$ for NGC~1333 and IC~348, respectively, which are consistent with the expected cosmic-ray energy density in the interstellar medium \citep{Ferriere2001}.

\begin{table*}[htbp]
\centering
\caption{Best-fitting parameters for the CR proton spectra of NGC~1333 and IC~348 given by Eq.~(\ref{eq:jparam}).}
\label{tab:params}
\begin{tabular}{lcccc}
\toprule\toprule
source & $C$ & $p_0$ [eV/c] & $\alpha$ & $\beta-\alpha$\\
\midrule
NGC 1333 & $(5.738\pm1.721)\times10^{-3}$ & $(12.90\pm3.472)\times10^{-2}$ & $-0.334\pm0.161$ & 2.822\\
IC 348   & $(10.47\pm3.141)\times10^{-3}$ & $(5.430\pm1.204)\times10^{-2}$ & $-0.595\pm0.102$ & 2.697\\
model $\mathscr{L}$ & $4.496\times10^{-3}$ & $8.826\times10^{-3}$ & 4.075 & 2.853\\
\bottomrule
\end{tabular}
\end{table*}

Finally, Fig.~\ref{fig:zvsN} shows a comparison between the model predictions and the observational estimates of $\zeta^{\rm ion}$ for each line of sight. 
Despite the rather large scatter in the observational estimates,  the models describe the decrease in $\zeta^{\rm ion}$ as the column density increases reasonably well. 
This may be better appreciated by comparing the data with the binned data. 

The factor of $\sim 10$ difference in CR ionization rates between IC 348 and NGC 1333 can be understood by evaluating the function $E\, d\zeta^{\rm ion}/dE$ 
at a column density $N(\mathrm{H}_2) \sim 10^{22}\,\mathrm{cm^{-2}}$, representative of these molecular clouds. 
At this column density, the function peaks at $\sim 10$ MeV, indicating that most of the ionization is produced by protons of this energy. 
The ratio of the two spectra in Fig.~\ref{fig:CRspectra} at $E=10$~MeV 
gives the observed factor of $\sim 10$.
Note that differences in the normalization of the Fermi spectra at energies above $\sim 3$~GeV have no impact on the ionization of these clouds, since the stopping range of a 3~GeV proton is $\sim 10^{26}\,\mathrm{cm^{-2}}$ \citep{Padovani2018a}. Consequently,
only at very large column densities, such as those typical of circumstellar disks, does the high-energy portion of the spectrum have an
influence on the value of $\zeta^{\rm ion}$. 
At lower column densities, the energy ranges $10$--$100$~MeV and $1$--$10$~GeV are effectively decoupled. 

\section{Discussion}
\label{sec:discussion}

\subsection{Bias of $n({\rm H}_2)$ and the impact on the CRIR}\label{sec:uncertainty of nh2}

Our derived gas volume density is based on the observed $^{13}$CO emission and therefore traces only the gas component in which CO remains in the gas phase. If a significant fraction of CO is frozen onto dust grains, then the true gas volume density could be underestimated, and CO would no longer be a fully reliable tracer of the total gas density.

However, we do not expect this effect to significantly alter our CRIR estimates, for two reasons. First, in the absence of internal particle acceleration sources, the bulk of the HINSA absorption is expected to arise from the more extended envelope rather than from the densest inner regions. Therefore, even if substantial CO depletion occurs in small, high-density regions (e.g., $n({\rm H}_2)\geq10^4$~cm$^{-3}$), their contribution to the observed HINSA feature should be limited. Second, any CO depletion is likely confined to relatively small spatial scales, so its impact should be diluted at our current angular resolution. We do find the average CO abundance decreases by a factor of $\sim$2 in NGC~1333 (from $1.8\times10^{-4}$ to $9.2\times10^{-5}$) for $N({\rm H}_2)\gtrsim10^{22}$~cm$^{-2}$. This suggests that strong CO depletion is not significant on the beam-averaged scales relevant to our analysis.

\begin{figure*}
\centering
\includegraphics[width=0.95\linewidth]{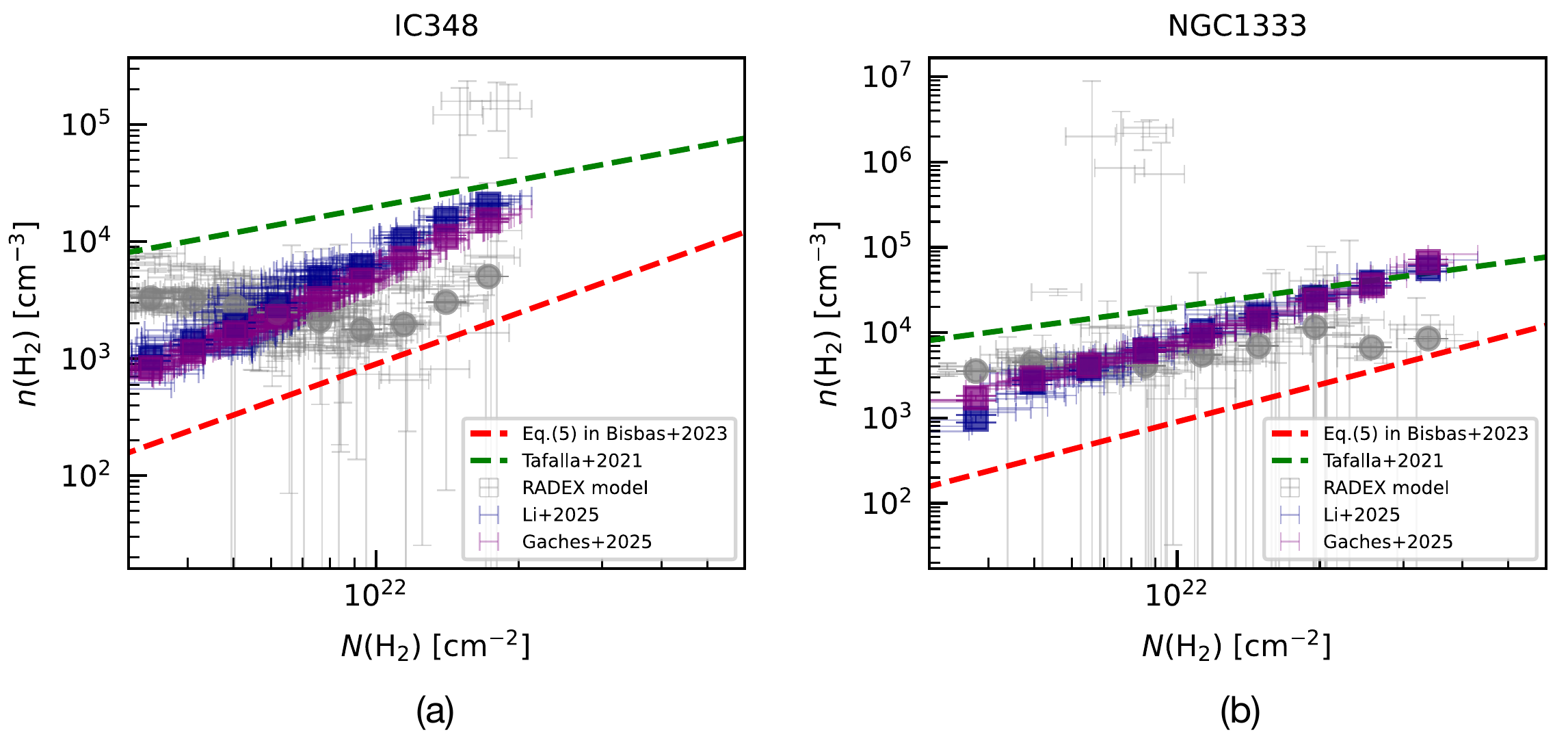}
\caption{{\it Panel~(a)\/}: estimates of $n({\rm H}_2)$ and weighted-average values in each bin from {\sc radex} modeling (gray), L25 (blue), and G25 (purple). The red and green dashed curves show the empirical relation from \citet{Bisbas2023} and the model used in \citet{Tafalla2021}. {\it Panel~(b)\/}: same as Panel~(a), for NGC~1333. \label{fig:figdensity}}
\end{figure*}

Figure \ref{fig:figdensity} shows the estimated $n({\rm H}_2)$ as a function of $N({\rm H}_2)$ in the two regions. For comparison, we also include the estimates of $n({\rm H}_2)$ from four recent approaches: ({\it i}\/) a three-dimensional (3D) density reconstruction algorithm from {\it Herschel} dust continuum \citep[hereafter, L25,][]{Li2025}; ({\it ii}\/) an empirical relation from hydrodynamical simulations \citep[hereafter, B23,][]{Bisbas2023}; ({\it iii}\/) the probabilistic model from {\it Herschel} dust continuum\footnote{The model decomposes the observed line-of-sight column density into a diffuse ``turbulent'' component and a dense ``gravitational'' component, we take the ``gravitational'' density component for comparison.} \citep[hereafter, G25,][]{Gaches2025b}; and ({\it iv}\/) the density model used for reconstructing line emissions of dense gas tracers in Perseus \citep[hereafter, T21,][]{Tafalla2021}. Although all these models suggest a power-law relationship between $n({\rm H}_2)$ and $N({\rm H}_2)$ (with different power indices), the densities estimated by different models can differ by more than an order of magnitude. The density function in T21 is much higher than the others since it is based on dense gas tracers (HCN, CS) and probes predominantly dense cores. 

If we compared the densities derived from $^{13}$CO transitions with these power-law relations, we find the density function is very similar to L25 and G25 in NGC~1333 (except for the highest $N({\rm H}_2)$ regions), while the density function in IC~348 shows a ``flat'' trend at $N({\rm H}_2)\leq10^{22}$~cm$^{-2}$. Such a ``flat'' density behavior was also found by previous PDR modelings using different molecules across the boundary of IC~348 \citep{Luo2023a}. The distinct $n({\rm H}_2)$--$N({\rm H}_2)$ relations in IC~348 and NGC~1333 are consistent with the differences in their column-density PDFs: IC~348 shows a broader lognormal component and a steeper high-density power-law tail than NGC~1333 \citep{Pezzuto2021}, suggesting different cloud structures. At high column densities ($N({\rm H}_2)\geq10^{22}$~cm$^{-2}$), the estimates from $^{13}$CO are 2--5 times smaller than L25 and T21. This is not surprising since dust emission traces the total column density and can be dominated by compact high-density clumps, while $^{13}$CO emission traces mostly envelopes. 

\begin{figure}
\centering
\includegraphics[width=0.95\linewidth]{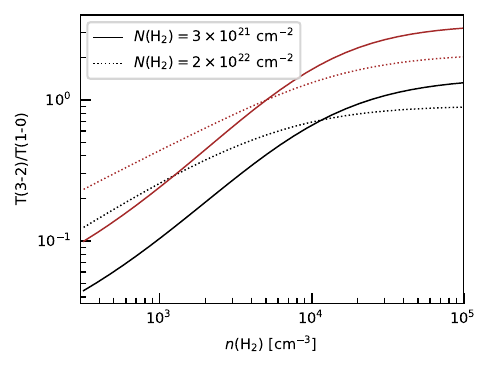}
\caption{Intensity ratio of $^{13}$CO 3--2/1--0 as a function of $n({\rm H}_2)$ from {\sc radex} modeling. The solid and dotted curves represent low and high column density, $N({\rm H}_2)=3$ and $20\times10^{21}$~cm$^{-2}$, respectively. The black and red curves represent $T_{\rm k}=15$ and 30~K, respectively. 
\label{fig:radex}}
\end{figure}

Figure \ref{fig:radex} shows the predicted intensity ratio of $^{13}$CO 3--2/1--0 at two different column densities (assuming $^{12}$C/$^{13}$C =65 and $^{12}$CO/H$_2=10^{-4}$) and temperatures, the ratio is sensitive to both density and temperature. However, the intensity ratio of $^{13}$CO 3--2/1--0 throughout the region exceeds 0.3, placing a lower limit on $n_{\rm H}=n({\rm H})+2n({\rm H}_2)$ of $1.5\times 10^3$~cm$^{-3}$. Thus, the low $n({\rm H}_2)$ values implied by B23 and L25/G25 have been ruled out. At high $N({\rm H}_2)$, the intensity ratio will not be sensitive to $n({\rm H}_2)$ above $10^4$~cm$^{-3}$, the estimates from {\sc radex} will induce large uncertainties. This may explain the sudden jumps of estimates in Fig.~\ref{fig:zvsN} and \ref{fig:figdensity}. On the other hand, as $N({\rm HINSA})$ is proportional to the line-of-sight integral of $\zeta^{\rm ion}$, it seems unlikely that a significant fraction of the HINSA column originates from such high-density regions, except for the existence of strong embedded CR acceleration sources. For this reason, we think the unexpected high-density jumps and therefore the derived high values of $\zeta^{\rm ion}$ for these data points are unreliable. Considering that the values of $T_{\rm k}$ may be underestimated at low $N({\rm H}_2)$ due to gas-dust decoupling, the resulting variance on $n({\rm H}_2)$ should remain within a factor of 2, even if $T_{\rm k}$ is underestimated by a factor of 2. 

We note that while the CRIR at the highest column densities would increase by up to a factor of a few if we use the $n({\rm H}_2)-N({\rm H}_2)$ power-law approaches (L25 and G25), it will not change the conclusion that IC~348 has a globally higher CRIR than that of NGC~1333.

\subsection{The interpretation of the observed disparity between IC~348 and NGC~1333}\label{sec:interpretation}

While the value of $\zeta^{\rm ion}$ derived from HINSA may be overestimated if a cloud lacks chemical equilibrium, IC~348 is evolutionarily more mature than NGC~1333 \citep{Hatchell2007,Bally2008,Young2015,Olivares2023}. Consequently, the elevated $\zeta^{\rm ion}$ observed in IC~348 is unlikely to result from differences in evolutionary timescales, but rather from intrinsic physical properties. Below, we discuss three possible factors contributing to the observed disparity between these two clouds.

\subsubsection{Origin of LECRs}

While the estimated LECR spectrum suggests that LECRs in NGC~1333 may undergo substantial attenuation, the considerably higher CRIR in IC~348 points to additional local sources of LECRs, possibly from outside of the cloud. This finding is consistent with recent $\gamma$-ray observations \citep{Jiang2025}, which report a higher CR spectrum in IC~348 relative to the local interstellar spectrum, whereas NGC~1333 exhibits a deficit below $\sim$2~GeV. Taken together, these results imply that the LECR population may originate, at least in part, from the same sources responsible for the harder GeV $\gamma$-ray spectrum. The decreasing trend of the CRIR at $N({\rm H}_2)\leq10^{22}$~cm$^{-2}$ agrees with previous chemical modeling of the IC~348 boundary \citep{Luo2023a}. 

Theoretical work indicates that LECRs can be efficiently accelerated at stellar-wind termination shocks in young, massive star clusters \citep{Casse1980,Blandford1987,Morlino2021,Menchiari2024}, potentially contributing a non-negligible fraction of the total cosmic-ray population \citep{Gupta2020,Peron2024}. Although both clusters host B-type stars, IC~348 lies in a region of lower column density and contains roughly twice as many members as NGC~1333 \citep[478 vs. 203,][]{Luhman2016}. By contrast, the NGC~1333 cluster is more deeply embedded. The resulting differences therefore provide a plausible explanation for the observed discrepancy in CRIR between the two clouds.  
Nevertheless, because the current spatial resolution and sensitivity limit the detection at high column densities, we cannot exclude the presence of internal acceleration sources within dense clumps. The embedded low-mass protostellar jet HH~211 inside IC~348 is probably not energetic enough to accelerate LECRs efficiently. Future high-resolution observations toward this region may give us more clues.

\subsubsection{Contribution from protostars}

Theoretical models suggest LECRs can be efficiently accelerated in protostellar jet shocks or accretion on the protostellar surface \citep{Padovani2015,Padovani2016,Gaches2018}. While previous CRIR mapping in NGC~1333 suggested enhancements extending several to tens of $10^4$~au around YSOs \citep{Pineda2024}, our observations did not observe a high CRIR on a larger scale ($>5\times10^4$~au). However, as the estimates of CRIR in the analytic method are proportional to $n({\rm H}_2)$, the uncertainties in $n({\rm H}_2)$ still leave the open question as to whether an enhancement of CRIR is directly related to the acceleration by protostellar objects. In addition, the two methods likely probe different gas components, as DCO$^+$ probe cold, dense cores while HINSA mainly originates from cold molecular envelopes. Comparing the results from IC~348 and NGC~1333, the discrepancy is notable given that NGC~1333 exhibits higher star-formation rate \citep[52 vs. 36~$M_\odot$~Myr$^{-1}$,][]{Young2015}\footnote{Assuming a mean stellar mass of 0.5~$M_\odot$.}, more than an order of magnitude higher outflow momentum (19.4 vs. 0.5 $M_\odot$~km~s$^{-1}$), and five time larger turbulent energy \citep[17 vs. 2.6 $\times 10^{37}$~J,][]{Curtis2010b} than that of IC~348. The result suggests that the small-scale CRIR enhancements from low-mass protostars in NGC~1333, if exists, will not contribute to the large-scale CRIR. 

\subsubsection{The effect of cloud structure}\label{sec:model}

The estimates of CRIR from HINSA may be biased by the complex cloud structure, as far-ultraviolet (FUV) photons can penetrate deeper and dissociate H$_2$ at higher projected $N({\rm H}_2)$. To evaluate this effect, we performed PDR simulations using the publicly available {\sc 3d-pdr} code\footnote{https://uclchem.github.io/3dpdr/} \citep{Bisbas2012}. 
We choose three uniform density clouds ($n_{\rm H}=10^3, 10^4, 10^5$~cm$^{-3}$) as control samples. The uniform density simulations are conducted in one-dimensional (1D) mode with {\sc 3d-pdr}. To investigate the effect of cloud structure, we constructed different density structures, in which the density field is controlled by the fractal dimension $\mathcal{D}$ \citep[2.0, 2.3, and 2.6 in our simulations, see][for more details]{Walch2012}. Higher fractal dimensions correspond to structures dominated by smaller, more homogeneously distributed clumps. 
The density probability distribution function ($n$-PDF) is normalized to 300~cm$^{-3}$ (allowing enough statistics in the column density range the same as in our observations) and cloud physical size is set to 3~pc (comparable to observations). Each cloud has a total number of 256$^3$ cells. Figure \ref{fig:FigB1} shows the projected total hydrogen column density $N_{\rm H}$ for three different clouds. 

To reduce computational resources, we use a reduced network containing 33 species and 330 reactions from the UMIST2012 network \citep{McElroy2013}. The initial abundances are adopted the same as in \citet{Luo2023a}. The FUV (isotropic) intensity is set as $\chi/\chi_0=100$ (normalized to the spectral shape given in \citet{Draine1978}) according to the estimates from total infrared flux \citep{Luo2023a}, and the CRIR is set as a constant ($\zeta_0 = 2\times10^{-17}$~s$^{-1}$) for all models.
All simulations were executed using the identical physical and chemical prescriptions. 

\begin{figure}
\centering
\includegraphics[width=0.95\linewidth]{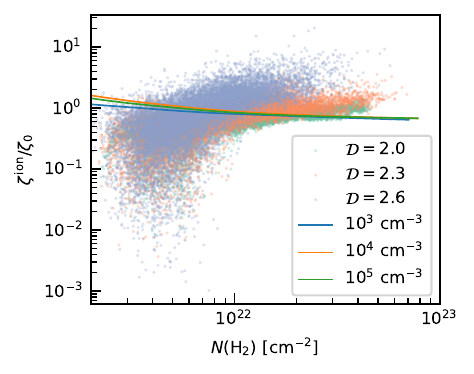}
\caption{Comparison between {\sc 3d-pdr} simulations and uniform density ($n_{\rm H}$ as labeled) simulations, showing the ratio of derived $\zeta^{\rm ion}$ from Eq.~(\ref{eq:zeta_ion}) and the model input $\zeta_0$ as a function of $N({\rm H}_2)$. The data points are colored by mass-weighted average densities along the HINSA sightlines. \label{fig:3DPDR_zeta}}
\end{figure}

We obtain the $N({\rm HINSA})$, $N({\rm H}_2)$, and $N_{\rm H}$ maps (see Appendix \ref{sec:maps}) from {\sc 3d-pdr} simulations to derive the values of $\zeta^{\rm ion}$ through Eq.~\ref{eq:zeta_ion}.  
We compare these results with the 1D simulations and among different 3D cloud models in Fig.~\ref{fig:3DPDR_zeta}. A notable feature is that, while the calculation through Eq.~\ref{eq:zeta_ion} in 1D simulations agree reasonably well with the model input over a wide range of $N({\rm H}_2)$, the calculated CRIR tends to be underestimated at lower projected $N({\rm H}_2)$ and overestimated at higher projected $N({\rm H}_2)$ in 3D simulations. At low $N({\rm H}_2)$, the gas is strongly affected by FUV and is difficult to cool down below 15~K to produce the HINSA feature (temperature threshold which we defined as HINSA in simulations, see Appendix \ref{sec:maps}), while a large fraction of cells can be molecular as long as the density is sufficiently high. At high $N({\rm H}_2)$, additional contributions arise from UV-driven dissociation. This effect is particularly important in clouds with high fractal dimension ($\mathcal{D}=2.6$), the estimates from HINSA are more overestimated because FUV radiation can penetrate to higher projected column densities. Although cloud structure could bias the inferred values, it is less likely that the cloud structure alone can explain an order-of-magnitude difference. 

\section{Conclusion}
\label{sec:conclusion}

Our new HINSA-based estimates of CRIR reveal a significant difference between two nearby low-mass star-forming clouds, IC~348 and NGC~1333, suggesting different CR acceleration sources for the two clouds. This again strengthens the environmental dependence of the CRIR. The low values of $\zeta^{\rm ion}$ in NGC~1333 imply that localized elevations of the CRIR near protostars, if present, do not overwhelm the ionization by Galactic CRs on larger scales. In contrast, the CRIR in IC~348 is an order of magnitude higher than NGC~1333, which aligns with $\gamma$-ray observations \citep{Yang2023,Jiang2025}. This may point to a common external acceleration mechanism responsible for both the enhanced ionization and the GeV $\gamma$-ray excess. Such a difference should have a large impact on CR-induced H$_2$ excitation in cold, dense cores, which might be detected by the {\it James Webb} Space Telescope (JWST) \citep{Bialy2020,Padovani+2022}.

A globally high CRIR in IC~348 implies a higher ionization fraction. This increases the coupling between gas and magnetic fields, thereby enhancing magnetic braking during protostellar core collapse. This process efficiently removes angular momentum, preventing the formation of large rotationally supported disks \citep{Kuffmeier2020}. Observational evidence supports this scenario: a recent survey of Class~0 and Class~I protostars in Perseus found that disks in IC~348 are, on average, smaller (18$\pm$7~au) than those in NGC~1333 \citep[31$\pm$11~au;][]{Segura-Cox2018}. 

Nonetheless, given the uncertainty regarding the variable gas density where HINSA exists and the fragmented nature of the molecular cloud, the absolute value of CRIR estimated from HINSA should be treated with caution, especially for distant targets where cloud structures may be blended within the synthesized beam. Although chemical modeling of fractal clouds indeed shows an impact on the HINSA-based estimates of $\zeta^{\rm ion}$, it is unlikely to produce an order-of-magnitude discrepancy seen in observations. We highlight the need for future detailed {\sc 3d-pdr} modeling with reconstructed cloud structures to better understand how cloud structure affects the ionization rate inferred from HINSA, as well as from other molecular line tracers.

\begin{acknowledgements}
We thank the anonymous referee for careful review, which improves the quality of the manuscript. We thank Jaime E. Pineda and Guang-xing Li for the helpful discussions, and Jaime E. Pineda for providing the NGC 1333 data cube for comparison. This work is  supported by NSFC grant No. 12588202. MP acknowledges the INAF-Minigrant 2024 ENERGIA (``ExploriNg low-Energy cosmic Rays throuGh theoretical InvestigAtions at INAF'');
DG acknowledges the INAF-Minigrant 2023 PACIFISM (``PArtiCles, Ionization and Fields in the InterStellar Medium''). 
TGB acknowledges support from the Leading Innovation and Entrepreneurship Team of Zhejiang Province of China (Grant No. 2023R01008).
BALG is supported by the German Research Foundation (DFG) in the form of an Emmy Noether Research Group - DFG project \#542802847. 
DL acknowledges support from the New Cornerstone foundation. 
This work has used the data from the Five-hundred-meter Aperture Spherical radio Telescope (FAST). FAST is a Chinese national mega-science facility, operated by the National Astronomical Observatories of Chinese Academy of Sciences (NAOC).
This publication utilizes data from Galactic ALFA HI (GALFA HI) survey data set obtained with the Arecibo L-band Feed Array (ALFA) on the Arecibo 305m telescope. The Arecibo Observatory is operated by SRI International under a cooperative agreement with the National Science Foundation (AST-1100968), and in alliance with Ana G. Méndez-Universidad Metropolitana, and the Universities Space Research Association. The GALFA HI surveys have been funded by the NSF through grants to Columbia University, the University of Wisconsin, and the University of California.
The James Clerk Maxwell Telescope has historically been operated by the Joint Astronomy Centre on behalf of the Science and Technology Facilities Council of the United Kingdom, the National Research Council of Canada and the Netherlands Organisation for Scientific Research.
This research has made use of data from the Herschel Gould Belt survey (HGBS) project (http://gouldbelt-herschel.cea.fr). The HGBS is a Herschel Key Programme jointly carried out by SPIRE Specialist Astronomy Group 3 (SAG 3), scientists of several institutes in the PACS Consortium (CEA Saclay, INAF-IFSI Rome and INAF-Arcetri, KU Leuven, MPIA Heidelberg), and scientists of the Herschel Science Center (HSC).
\end{acknowledgements}

% WARNING
%-------------------------------------------------------------------
% Please note that we have included the references to the file aa.dem in
% order to compile it, but we ask you to:
%
% - use BibTeX with the regular commands:
  % \bibliographystyle{aa} % style aa.bst
  % \bibliography{Yourfile} % your references Yourfile.bib
%
% - join the .bib files when you upload your source files
%-------------------------------------------------------------------
% \begin{thebibliography}

\bibliographystyle{aa}
\bibliography{reference} % if your bibtex file is called example.bib

% \end{thebibliography}

% \end{document}

\begin{appendix} %First  appendix

\section{An example of HINSA fitting.}
\label{sec:hinsa_fit}

\begin{figure}[H]
\centering
\includegraphics[width=0.95\linewidth]{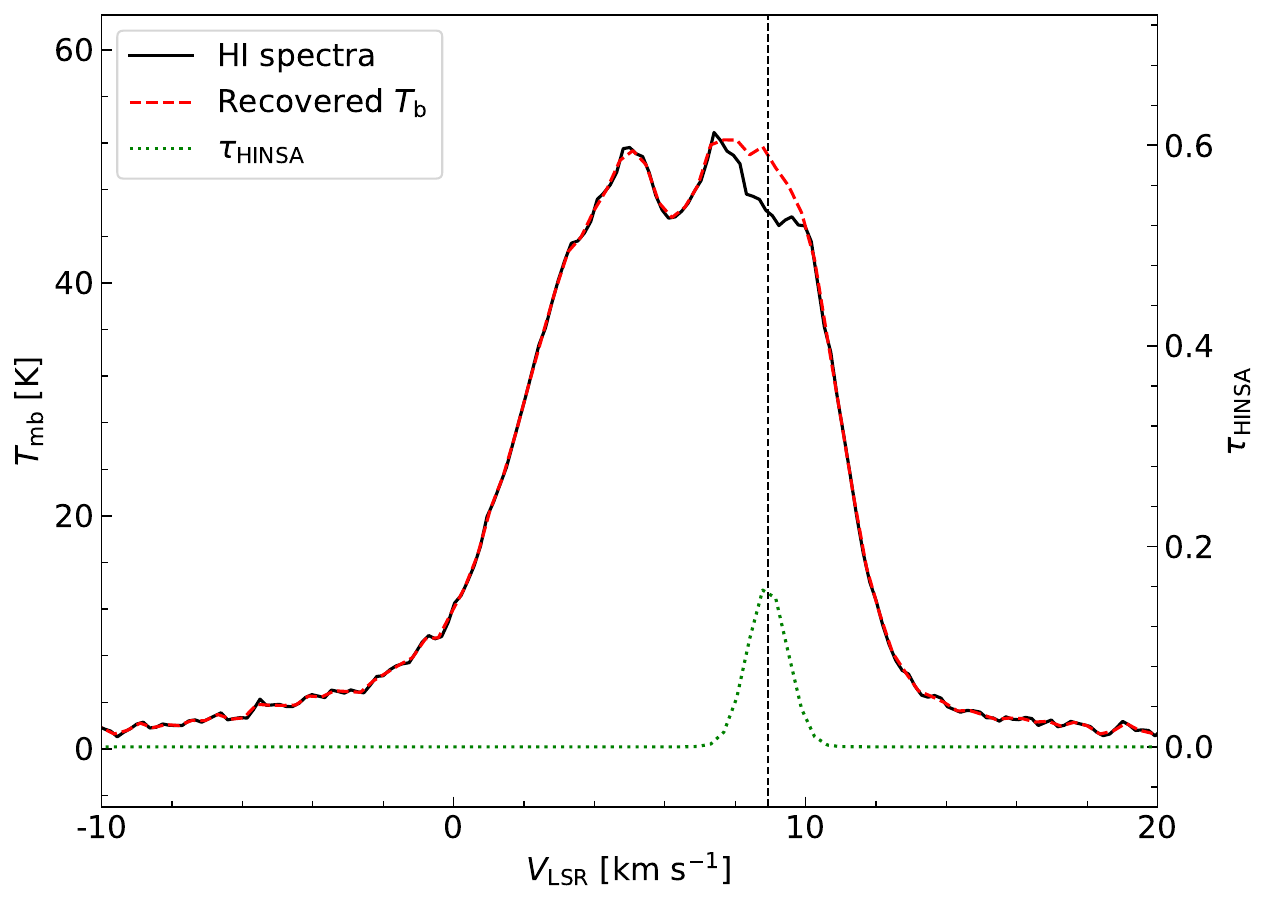}
\caption{Example shows the observed H\,{\sc i} spectra (black solid curve), the recovered background H\,{\sc i} emission without HINSA absorption (red dashed curve), and decomposed optical depth $\tau$ (green dotted curve). The black vertical lines denote the decomposed $V_{\rm lsr}$ of HINSA. \label{fig:hinsa_fit}}
\end{figure}

Figure~\ref{fig:hinsa_fit} shows an example of HINSA fitting from the center of IC~348. The continuum background (cosmic microwave background + synchrotron emission) in the fitting procedure is set to be 3.3~K, and the assumed fraction of H\,{\sc i} in the foreground is 0.9 when we recover the unabsorbed spectrum \citep[see, e.g.,][]{Luo2024a,Luo2025}. We consider a 20\% uncertainty in addition to the HINSA fitting due to these assumptions.

\section{$N_{\rm H}$, $N({\rm HINSA})$, $n_{\rm H}$, and $N({\rm H}_2)$ maps from {sc 3d-pdr} simulations.}\label{sec:maps}

\begin{figure*}
\centering
\includegraphics[width=0.95\linewidth]{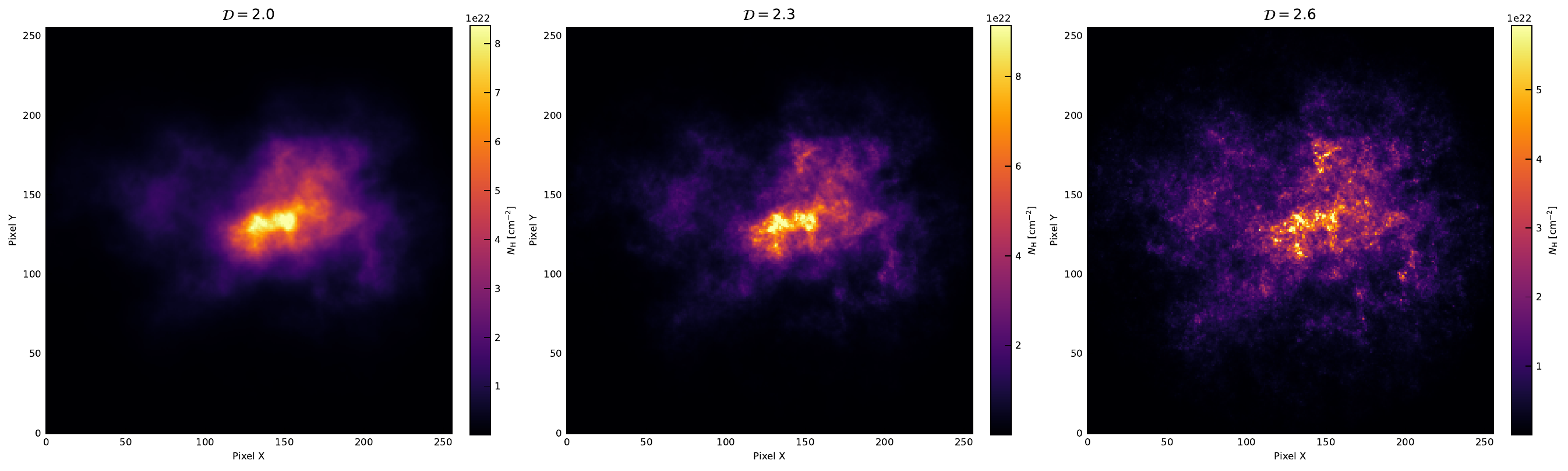}
\caption{Total H column densities ($N_{\rm H}$) from three clouds with fractal dimension $\mathcal{D}$ = 2.0 (left), 2.3 (middle), and 2.6 (right).
\label{fig:FigB1}}
\end{figure*}

\begin{figure*}
\centering
\includegraphics[width=0.95\linewidth]{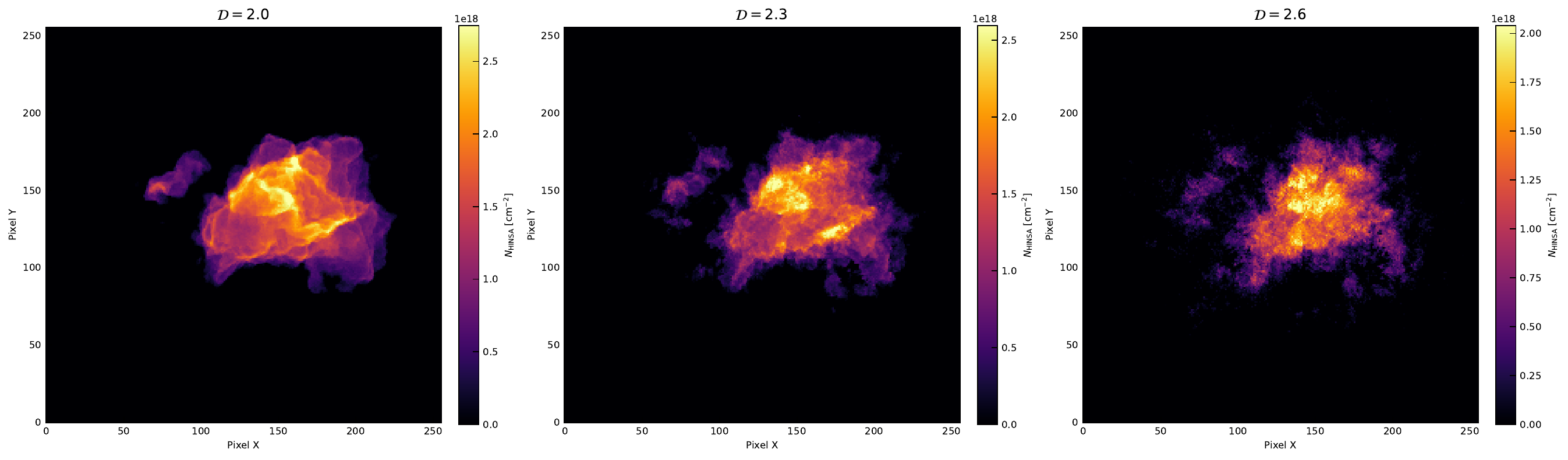}
\caption{The same as Fig.~\ref{fig:FigB1} but for $N({\rm HINSA})$.
\label{fig:FigB2}}
\end{figure*}

\begin{figure*}
\centering
\includegraphics[width=0.95\linewidth]{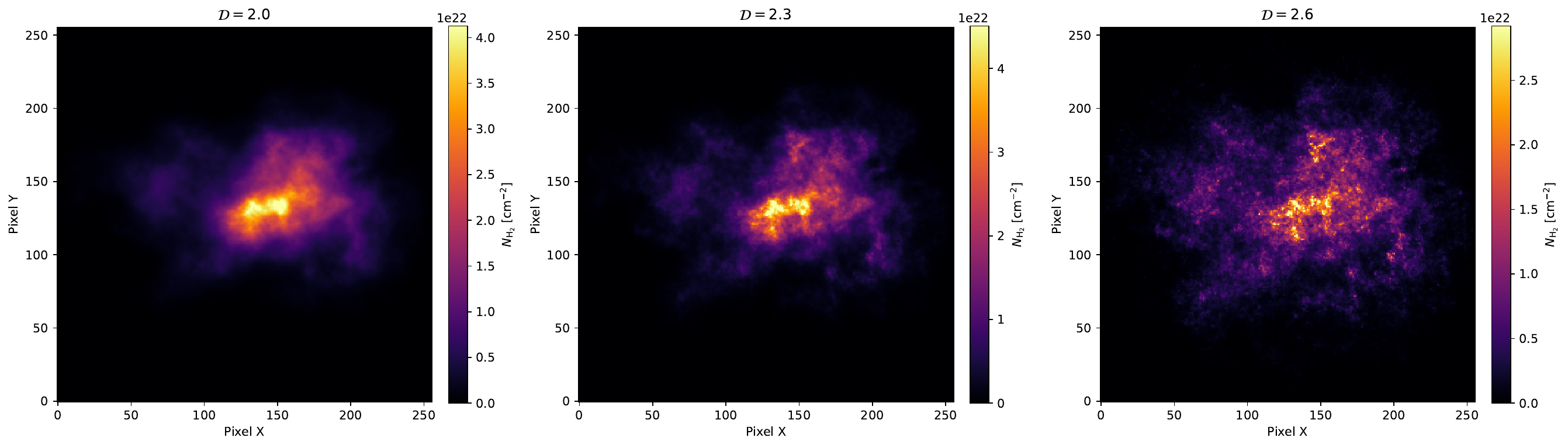}
\caption{The same as Fig.~\ref{fig:FigB1} but for $N({\rm H}_2)$.
\label{fig:FigB3}}
\end{figure*}

\begin{figure*}
\centering
\includegraphics[width=0.95\linewidth]{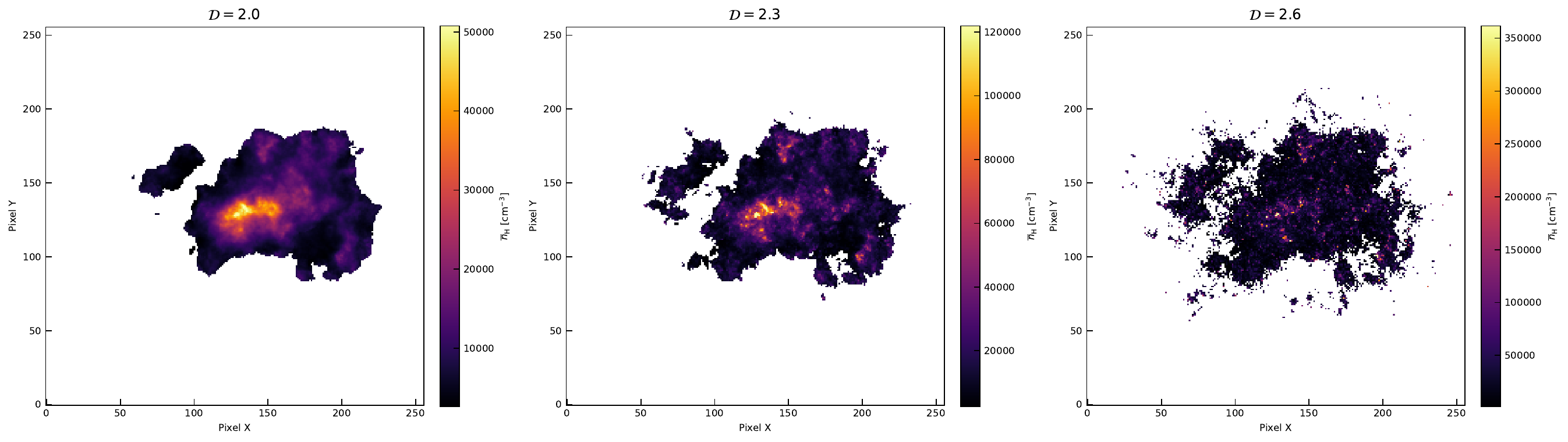}
\caption{The same as Fig.~\ref{fig:FigB1} but for $n_{\rm H}$.
\label{fig:FigB4}}
\end{figure*}

Figures \ref{fig:FigB1} to \ref{fig:FigB4} show the $N_{\rm H}$, $N({\rm HINSA})$, $n_{\rm H}$, and $N({\rm H}_2)$ maps from {\sc 3d-pdr} simulations. The HINSA column density along each line of sight was computed by summing the H\,{\sc i} from cells with gas temperature $T_{\rm gas}<15$~K. We note that the temperature criterion is only an approximation, and the temperature of ``true'' HINSA cells is not to be one fixed value, but these cells should be cold and exist only in molecular-dominated regions. A different definition of $T_{\rm gas}$ may give a slightly different column density of HINSA. 

\section{Dissociation of H$_2$ by cosmic rays}
\label{revised_zetah}

Direct cosmic-ray dissociation, dissociative ionization and, for cosmic-ray nuclei, dissociative electron capture dissociate H$_2$ at a rate
\begin{equation}
\zeta^{\rm diss}=4\pi\eta\left[\int_0^\infty j_p(E)\sigma_p^{\rm tot}(E)\, dE
+\int_0^\infty j_{\rm sec}(\varepsilon)\sigma_e^{\rm tot}(\varepsilon)\, d\varepsilon\right],
\label{zh_def}
\end{equation}
where $j_p(E)$ and $j_{\rm sec}(\varepsilon)$ are the specific intensities of cosmic ray
protons of energy $E$ and secondary electrons of energy $\varepsilon$, respectively, 
$\eta$ is a factor that accounts for helium and heavier cosmic-ray nuclei, $\sigma_p^{\rm tot}(E)$
and $\sigma_e^{\rm tot}(\varepsilon)$ are the total dissociation cross sections by
proton- and electron-impact on H$_2$, respectively.

\begin{figure*}
\centering
\includegraphics[width=0.4\linewidth]{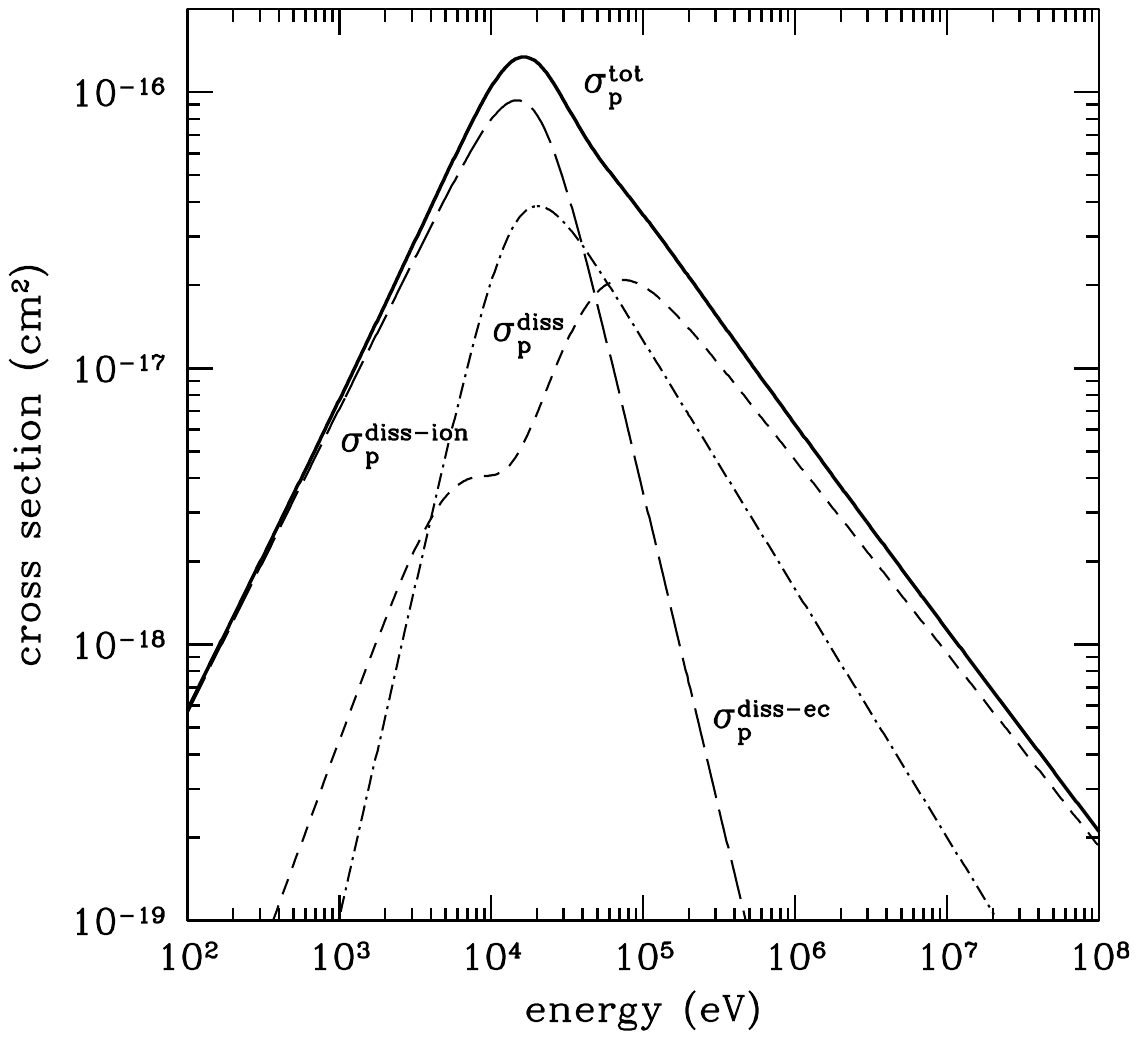}
\includegraphics[width=0.4\linewidth]{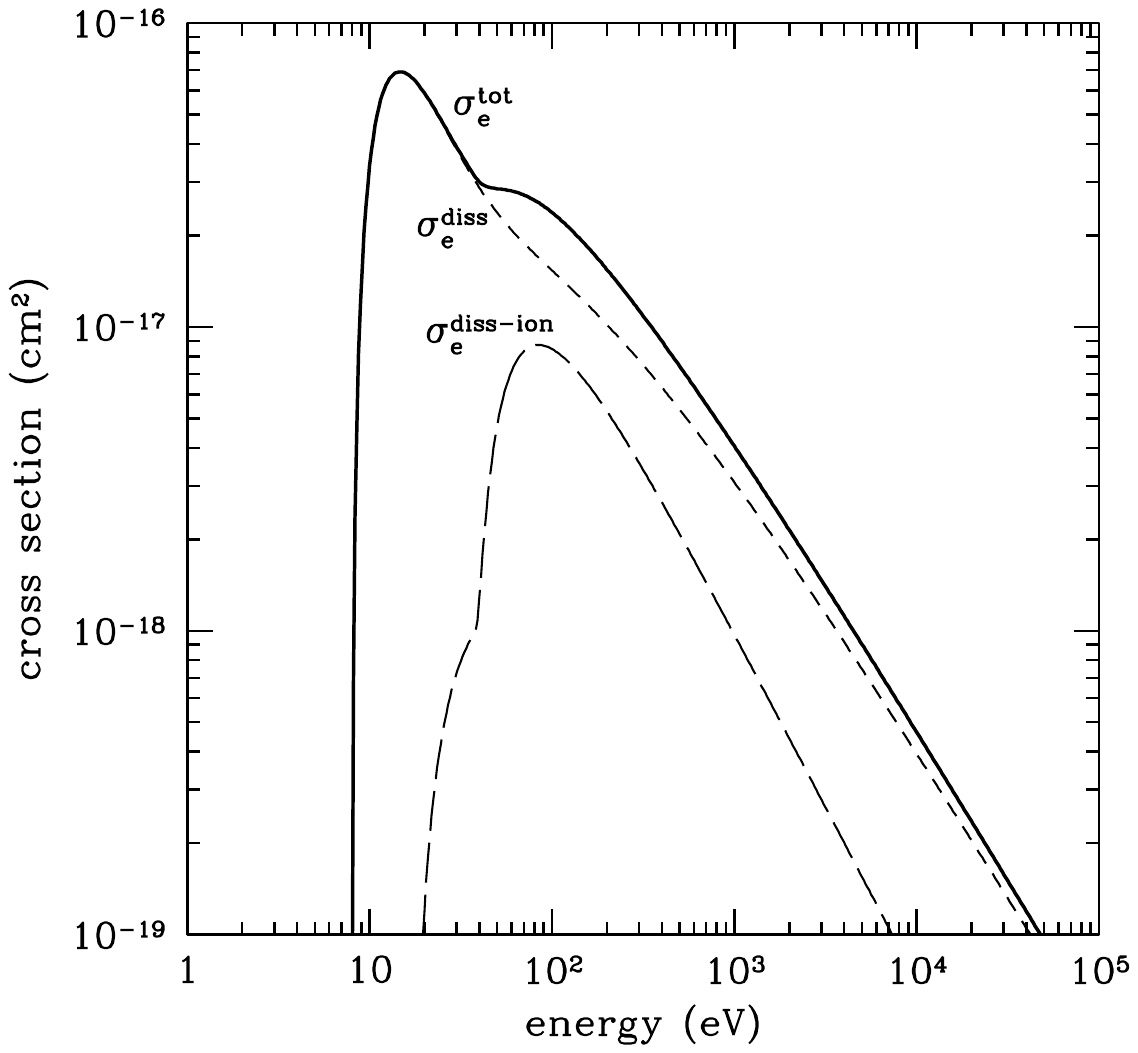}
\vspace{-60pt}
\caption{{\it Left panel:\/} proton-impact dissociation cross sections of H$_2$:
dissociation $\sigma_{p}^{\rm diss}$ \citep{Plowman26}; dissociative ionization $\sigma_{p}^{\rm diss-ion}$ \citep{Martinez03}; dissociative electron capture $\sigma_{p}^{\rm diss-ec}$ \citep{Martinez03}. The 
thick curve shows the total dissociation cross section.
{\it Right panel:\/} same, for electrons: dissociation $\sigma_{e}^{\rm diss}$ \citep{Scarlett19};
dissociative ionization $\sigma_{e}^{\rm diss-ion}$ \citep{Wunderlich11,Wunderlich16}. . 
\label{fig:Fig_cs_diss}}
\end{figure*}

Specifically, direct proton-impact processes that dissociate H$_2$ are: ({\it i}\/) dissociation 
\begin{equation}
    p + {\rm H}_2 \rightarrow p + 2{\rm H},
\end{equation}
with cross section $\sigma_p^{\rm diss}(E)$; 
({\it ii}) dissociative ionization
\begin{equation}
    p + {\rm H}_2 \rightarrow p+ {\rm H} + {\rm H}^+ + e,
\end{equation}
with cross section $\sigma_p^{\rm diss-ion}(E)$; and ({\it iii}) dissociative electron capture
\begin{equation}
    p + {\rm H}_2 \rightarrow {\rm H}_{\rm fast} + {\rm H}^+ + {\rm H},
\end{equation}
with cross section $\sigma_p^{\rm diss-ec}(E)$. Therefore,
\begin{equation}
    \sigma_p^{\rm H}(E)=\sigma_p^{\rm diss}(E)+\sigma_p^{\rm diss-ion}(E)+\sigma_p^{\rm diss-ec}(E)
\end{equation}
Similarly, electron-impact processes that produce H atoms are:  dissociation
\begin{equation}
    e + {\rm H}_2 \rightarrow e + 2{\rm H},
\end{equation}
with cross section $\sigma_e^{\rm diss}(\varepsilon)$; and  dissociative ionization
\begin{equation}
    e + {\rm H}_2 \rightarrow e + {\rm H} + {\rm H}^+ + e,
\end{equation}
with cross section $\sigma_e^{\rm diss-ion}(\varepsilon)$. 
Therefore,
\begin{equation}
    \sigma_e^{\rm H}(\varepsilon)=\sigma_e^{\rm diss}(\varepsilon)+\sigma_e^{\rm diss-ion}(\varepsilon).
\end{equation}
The relevant proton- and electron-impact cross sections are shown in Fig.~\ref{fig:Fig_cs_diss}. 
The value 
of $\zeta^{\rm diss}/\zeta^{\rm ion}$, calculated with the cosmic-ray proton and electron spectra described in 
Sect.~4.2, depends very weakly on the cloud's column density and is equal to  
$0.589$ at $N({\rm H}_2)=10^{22}$~cm$^{-2}$, with secondary electrons contributing 5.5 times more than nuclei to the production of H atoms. This value must be considered more accurate than that computed earlier by \cite{Padovani2018b},
who neglected dissociative ionization and dissociative electron capture, and adopted an approximated procedure
for the calculation of the spectrum of secondary electrons.

\end{appendix}
\end{document}